\documentclass[11pt]{article}
\usepackage{hyperref}
\pdfoutput=1

\begin{document}

\title{To be, or not to be... 
\\-instabilities on a liquid jet penetrated into a flowing bath-}

\author{Kaoru Hattori$^{1}$, Ichiro Ueno$^{2}$ \\
\\\vspace{0pt} $^{1}$Undergraduate,Department of Mechanical Engineering, Tokyo University of Science, \\\vspace{6pt} 2641 Yamazaki, Noda, Chiba 278-8510, JAPAN
\\\vspace{6pt} $^{2}$Department of Mechanical Engineering, Tokyo University of Science, Noda, JAPAN}

\maketitle

%% The abstract (in this file, and that submitted as text to arXiv) shouldinclude the exact phrase
%% "fluid dynamics video" or "fluid dynamics videos"

\begin{abstract}
Dynamic behaviors of the penetrated jet and the departure of the bubble of wrapping gas at the tip of the collapsing jet observed by use of a high-speed camera are included in this fluid dynamics video. 

\end{abstract}
% main text

\section{Introduction}

%The {\em hyperref} package is used to make links to the videos.
%% The format is: \href{URL of video}{name that will appear in the text}

%Two videos are
%\href{http://ecommons.library.cornell.edu/bitstream/1813/8237/2/LIFTED_H2_EMS
%T_FUEL.mpg}{Video
%1} and
%\href{http://ecommons.library.cornell.edu/bitstream/1813/8237/4/LIFTED_H2_IEM
%_FUEL.mpg}{Video
%2}.

%It is recommended that the article include:
%\begin{enumerate}

We conduct a series of experiments with a special interest on a penetration process and instabilities arisen on a liquid jet impinged to a liquid of the same kind flowing in a channel. The impinged jet penetrates into the flowing bath accompanying with entrainment of the ambient immiscible gas, which results in the impinged jet wrapped by the entrained gas as a 'sheath.' This sheath formation enables the impinged jet to survive in the fluid in the channel without coalescing until the entrained-air sheath breaks down. Occasionally a 'cap' of the entrained air is formed at the tip of the penetrated jet, and the jet elongates like a long balloon.

%\item In the Abstract (in the LaTeX file and in the text submitted
%to arXiv), the exact phrase ``fluid dynamics video" or ``fluid
%dynamics videos". This is to facilitate subsequent searching.
%\end{enumerate}
%
\end{document}